\documentclass[english]{llncs}
\usepackage[T1]{fontenc}
\usepackage[latin9]{inputenc}
\usepackage{refstyle}
\usepackage{pmboxdraw}
\usepackage{amstext}

\makeatletter

\AtBeginDocument{\providecommand\secref[1]{\ref{sec:#1}}}
\providecommand{\tabularnewline}{\\}
\RS@ifundefined{subsecref}
  {\newref{subsec}{name = \RSsectxt}}
  {}
\RS@ifundefined{thmref}
  {\def\RSthmtxt{theorem~}\newref{thm}{name = \RSthmtxt}}
  {}
\RS@ifundefined{lemref}
  {\def\RSlemtxt{lemma~}\newref{lem}{name = \RSlemtxt}}
  {}

\makeatother

\usepackage{babel}
\usepackage{amsmath}
\usepackage{amsfonts}
\usepackage{amssymb}

\begin{document}

\title{Fully Mechanized Proofs of Dilworth's Theorem and Mirsky's Theorem }

\author{Abhishek Kr.\ Singh\thanks{Email:~abhishek.uor@gmail.com;~
		~URL:~http://www.tcs.tifr.res.in/\~\/abhishek} }

\institute{School of Technology \& Computer Science\\
	Tata Institute of Fundamental Research\\
	Mumbai 400005, INDIA}

\date{01/09/16}

\maketitle

\date{07/15/16}
\begin{abstract}
We present two fully mechanized proofs of Dilworth's and Mirsky's
theorems in the Coq proof assistant. Dilworth's Theorem states that
in any finite partially ordered set (poset), the size of a smallest
chain cover and a largest antichain are the same. Mirsky's Theorem
is a dual of Dilworth's Theorem. We formalize the proofs by Perles
\cite{key-2} (for Dilworth's Theorem) and by Mirsky \cite{key-5}
(for the dual theorem). We also come up with a library of definitions
and facts that can be used as a framework for formalizing other theorems
on finite posets.
\end{abstract}

\section{Introduction}

Formalization of any mathematical theory requires a lot of time and effort. The
length of formal proofs blow up significantly. In combinatorics, the
task becomes even more difficult due to the lack of structure in the
theory. There is no established order to formalize the theory. Some statements often admit more than one proof using completely
different ideas. Thus, exploring the dependencies among the results
may help in identifying an effective order to formalize them. Dilworth's
Theorem is a well-known result in combinatorics. It relates the size
of a chain cover and an antichain in a poset. The original version,
which talks about the chain cover, was first proved by Dilworth \cite{key-1}.
Since then, the theorem has attracted significant attention and several
new proofs \cite{key-2,key-3,key-4} were found. Besides being popular,
Dilworth's Theorem is an important result as it reveals the structure
of a general poset. It has been successfully used to give intuitive
and concise proofs of some other important results such as Hall's
Theorem \cite{key-6-1,key-6-2}, Erd\H{o}s-Szekeres Theorem \cite{key-6-4}
and Sperner's Lemma \cite{key-6-5,key-6-3}. In this sense, it is
a central theorem and a good candidate for formalization. For Dilworth's
Theorem we have formalized the proof by Perles \cite{key-2}. We also
formalize a dual of the Dilworth's Theorem (Mirsky's Theorem \cite{key-5})
which relates the size of an antichain cover and a chain in a poset. 

Formalization of the known mathematical results can be traced back
to the systems Automath and Mizar \cite{key-13}. Mizar hosts the
largest repository of the formalized mathematics. It uses a declarative
proof style close to the language that mathematicians understand.
Mizar supports some built in automation to save time during proof
development. However, this results in a large kernel (core) and reduces
our faith in the system. The Coq proof assistant deals with this problem
in a nice way. It separates the process of proof development from
proof checking. Some small scale proof automation is also possible
in Coq. However, every proof process finally yields a proof-term which
is verified using a small kernel. Thus, the part (kernel) of the code
we need to trust remains small. 

In addition to a small kernel, the Coq proof assistant also has some
other distinctive features that we found useful: 
\begin{description}
\item [{\emph{Records}}] \emph{(Dependent):} It helps us to pack mathematical
objects and their properties in one definition. For example, in the
Coq standard library different components of a partial order and their
properties are expressed using a single definition of dependent record
(PO) \cite{key-6}. 
\item [{\emph{Coercions}:}] It helps us in defining hierarchy among mathematical
structure. For example, in our formalization finite partial orders
are defined by extending the definition of partial orders with finiteness
condition. This avoids redefining similar things at different places.
It is similar to the concept of inheritance in programming languages. 
\item [{\emph{Ltac}:}] A language support \cite{key-8} for creating new
tactics while being completely inside Coq. It helps in small scale
proof automation and proof search, hence reduces the time of proof
development significantly. 
\item [{\emph{Library}:}] The Coq system hosts a standard library \cite{key-14}
that contains a bunch of useful definitions and results. We use this
facility and avoid new definitions, unless absolutely essential. For
example, we use the modules for Sets and Basic Peano arithmetic extensively
in this work. 
\end{description}
In this paper, we give the details of our mechanized proofs of Dilworth's
and Mirsky's Theorems. In \secref{Dilworth-and-Dual-Dilworth} we
present the original theorems and their proofs by Perles \cite{key-2}
and Mirsky \cite{key-5}. In \secref{Definitions-and-Formalization}
we describe every term that appears in the formal statements of these
theorems. We provide a detailed account of some general results on
finite partial orders in \secref{Some-results-on-FPO}. Finally, we
review the related works in \secref{Related-Work} and conclude in
\secref{Conclusions}. 

\section{Dilworth's and Mirsky's Theorem\label{sec:Dilworth-and-Dual-Dilworth}}

\subsection{Partially ordered sets (poset)}

A poset $(P,\leq)$ consists of a set P together with a binary relation
$\leq$ satisfying reflexivity, antisymmetry and transitivity properties.
Elements $a,\,b\in P$ are said to be \emph{comparable }if $a\leq b$
or $b\leq a$. Otherwise, they are \emph{incomparable}. A \emph{chain}
is a subset of P any two of whose elements are comparable. A subset
of P in which no two distinct elements are comparable is called an
\emph{antichain}. Note that 
\begin{itemize}
\item A chain and an antichain can have at most one element in common. 
\end{itemize}
A \emph{chain cover} is a collection of chains such that their union
is the entire poset. Similarly, an \emph{antichain cover }is a collection
of antichains such that their union is the entire poset. 
\begin{itemize}
\item The \emph{width} of a poset P, $width(P)$, is the size of a largest
antichain in P. 
\item The \emph{height} of a poset, $height(P),$ is the size of a largest
chain in P. 
\end{itemize}
An element $b\in P$ is called a \emph{maximal element }if there is
no $a\in P$ (different from $b$) such that $b\leq a$. Similarly,
an element $a\in P$ is called a \emph{minimal element} if there is
no $b\in P$ (different from $a$) such that $b\leq a$. 

\subsection{Dilworth's Theorem }

\textbf{Theorem} (Dilworth): In any poset, the maximum size of an
antichain is equal to the minimum number of chains in any chain cover.
In other words, if $c(P)$ represents the size of a smallest chain
cover of P, then $width(P)=c(P)$.\\
 \\
\textbf{Proof} (Perles): The equality will follow if one can prove:
\begin{enumerate}
\item Size of an antichain $\leq$ Size of a chain cover, and
\item There is a chain cover of size equal to $width(P)$.
\end{enumerate}
It is easy to see why (1) is true. Assume otherwise, i.e, there is
an antichain $\mathcal{A}$ of size bigger than the size of a smallest
chain cover $\mathcal{C_{V}}$. Then $\mathcal{A}$ will have more
elements than the number of chains in $\mathcal{C_{V}}.$ Hence, there
must exist a chain $\mathcal{C}$ in $\mathcal{C_{V}}$ which covers
two element of $\mathcal{A}$. However, this cannot be true since
a chain and an antichain (in this case $\mathcal{C}$ and $\mathcal{A}$)
can have at most one element in common. 

Proof of (2) is more involved. We will prove (2) using strong induction
on the size of $P$. Let $m$ be the size of the largest antichain
in P, i.e, $m=width(P)$. 
\begin{itemize}
\item Induction hypothesis: For all posets $P'$ of size at most $n$, there
exists a chain cover of size equal to $width(P')$.
\end{itemize}
Induction step: Fix a poset P of size at most $n+1$. Let maximal(P)
and minimal(P) represents respectively the set of all maximal and
the set of all minimal elements of P. Now, one of the following two
cases might occur,
\begin{enumerate}
\item There exists an antichain $\mathcal{A}$ of size $m$ which is neither
maximal(P) nor minimal(P).
\item No antichain other than maximal(P) or minimal(P) has size $m$. 
\end{enumerate}
\textbf{Case-1:} For the first case we define the sets $P^{+}$ and
$P^{-}$ as follows:
\begin{center}
\begin{tabular}{c}
\tabularnewline
$P^{+}=\left\{ x\in P:\,x\geq y\,\,\text{for some }y\in\mathcal{A}\right\} $\tabularnewline
$P^{-}=\left\{ x\in P:\,x\leq y\,\,\text{for some }y\in\mathcal{A}\right\} $\tabularnewline
\tabularnewline
\end{tabular}
\par\end{center}

Here $P^{+}$ captures the notion of being above $\mathcal{A}$ and
$P^{-}$ captures the notion of being below $\mathcal{A}$. Note that
the elements of $\mathcal{A}$ are both above and below $\mathcal{A}$,
i.e, $\mathcal{A}\subseteq P^{+}\cap P^{-}$. For any arbitrary element
$x\in P$
\begin{itemize}
\item If $x\in A$ then $x\in P^{+}\cap P^{-}$ and hence $x\in P^{+}\cup P^{-}$.
\item If $x\notin\mathcal{A}$ then $x$ must be comparable to some element
in $\mathcal{A}$; otherwise $\{x\}\cup\mathcal{A}$ will be an antichain
of size $m+1$. Hence, if $x\notin\mathcal{A}$ then $x\in P^{+}\cup P^{-}$. 
\end{itemize}
Therefore, $P^{+}\cup P^{-}=P$. Since there is at least one minimal
element not in $\mathcal{A}$, $P^{+}\neq P$. Similarly $P^{-}\neq P$.
Thus $|P^{+}|<|P|$ and $|P^{-}|<|P|$, hence we will be able to apply
induction hypothesis to them. Observe that $\mathcal{A}$ is also
a largest antichain in the poset restricted to $P^{+}$; because if
there was a larger one, it would have been larger in $P$ also. Therefore
by induction, 
\begin{itemize}
\item There exists a chain cover of size $m$ for $P^{+}$, say $P^{+}=\cup_{i=1}^{m}C_{i}$.
\item Similarly, there is a chain cover of size $m$ for $P^{-}$, say $P^{-}=\cup_{i=1}^{m}D_{i}$.
\end{itemize}
Elements of $\mathcal{A}$ are the minimal elements of the chains
$C_{i}$ and the maximal elements of the chains $D_{i}$. Therefore
we can join the chains $C_{i}$ and $D_{i}$ together in pairs to
form $m$ chains which form a chain cover for the original poset $P$.

\textbf{Case-2: }In this case we can't have an antichain of size $m$
which is different from both maximal(P) and minimal(P). Consider a
minimal element $x.$ Choose a maximal element $y$ such that $x\leq y$.
Such a $y$ always exists. Remove the chain $\{x,y\}$ from $P$ to
get the poset $P'$. Then $P'$ contains an antichain of size $m-1$.
Also note that $P'$ can't have an antichain of size $m$. 
\begin{itemize}
\item Because if there was an antichain of size $m$ in $P'$ , then that
would also be an antichain in $P$ which is different from both maximal(P)
and minimal(P), and hence we would have been in the first case (i.e,
Case-1). 
\end{itemize}
Hence by induction hypothesis we get a chain decomposition of $P'$
of size $m-1$. These chains, together with $\{x,y\}$, give a decomposition
of $P$ into $m$ chains. $\square$

\subsection{Mirsky's Theorem}

\textbf{Theorem} (Dual-Dilworth): In any poset, the maximum size of
a chain is equal to the minimum number of antichains in any antichain
cover. In other words, if $c(P)$ represents the size of a smallest
antichain cover of P, then $height(P)=c(P)$.\textbf{}\\
\textbf{}\\
\textbf{Proof} (Mirsky): The equality will follow if one can prove:
\begin{enumerate}
\item Size of a chain $\leq$ Size of an antichain cover, and
\item There is an antichain cover of size equal to $height(P)$. 
\end{enumerate}
Again, it is easy to see why (1) is true. Any chain shares at most
one element with each antichain from an antichain cover. Moreover,
every element of the chain must be covered by some antichain from
the antichain cover. Hence, the size of any chain is smaller than
or equal to the size of any antichain cover. 

We will prove (2) using strong induction on the size of the largest
chain of $P$. Let $m$ be the size of the largest chain in P, i.e,
$m=height(P)$. 
\begin{itemize}
\item Induction hypothesis: For all posets $P'$ of height at most $m-1$,
there exists an antichain cover of size equal to $height(P')$.
\end{itemize}
Induction Step: Let $M$ denotes the set of all maximal elements of
$P$, i.e, $M=\text{maximal(P)}$. Observe that $M$ is a non empty
antichain and shares an element with every largest chain of $P$.
Consider now the partially ordered set $(P-M,\leq)$. The length of
the largest chain in $P-M$ is at most $m-1$. On the other hand,
if the length of the largest chain in $P-M$ is less than $m-1$,
$M$ must contain two or more elements that are members of the same
chain, which is a contradiction. Hence, we conclude that the length
of largest chain in $P-M$ is $m-1$. Using induction hypothesis there
we get an antichain cover $\mathcal{A_{C}}$ of size $m-1$ for $P-M$.
Thus, we get an antichain cover $\mathcal{A_{C}}\cup\{M\}$ of size
$m$ for $P$. $\square$

\section{Definitions and Formalization\label{sec:Definitions-and-Formalization}}

\subsection{Definitions }

Once a statement is proved in Coq, the proof is certified without
having to go through the proof-script. However, one needs to verify
whether the statement being proved correctly represents the original
theorem. In this section we try to explain the definition of each
term that appears in the formal statement of the Dilworth's theorem. 

\subsubsection*{Definitions from Standard Library}

We have used the \emph{Sets }module from the Coq Standard Library
\cite{key-14} , where a declaration \texttt{S: Ensemble U} is used
to represent a set $S$. 
\begin{itemize}
\item Sets are treated as predicates, i.e, $x\in S$ iff \texttt{S x} is
provable. Set belongingness is written as \texttt{In S x} instead
of just writing \texttt{S x}. 
\end{itemize}
Empty set is defined as a predicate \texttt{Empty\_set} which is not
provable anywhere. \texttt{Singleton x} and \texttt{Couple x y} represents
the sets $\{x\}$ and \{$x,y\}$ respectively. Partial Orders are
defined as a dependent record type in the standard library. It has
four fields,\texttt{ }
\begin{description}
\item [{\texttt{Record}}] \texttt{PO (U : Type) : Type := Definition\_of\_PO}~\\
\texttt{ \{ Carrier\_of : Ensemble U; }~\\
\texttt{Rel\_of : Relation U;}~\\
\texttt{ PO\_cond1 : Inhabited U Carrier\_of;}~\\
\texttt{ PO\_cond2 : Order U Rel\_of \}.}
\end{description}
First two fields represents the carrier set $S$ and the binary relation
$\leq$ of a partially ordered set $(S,\leq)$. Third field \texttt{PO\_cond1}
is a proof that S is a non-empty set. Similarly, \texttt{PO\_cond2}
is a proof that $\leq$ is an order (i.e, reflexive, transitive and
antisymmetric). Using the coercion feature of Coq we extend this definition
to define finite partial orders as 
\begin{description}
\item [{\texttt{Record}}] \texttt{FPO (U : Type) : Type := Definition\_of\_FPO
}~\\
\texttt{\{ PO\_of :> PO U ; }~\\
\texttt{FPO\_cond : Finite \_ (Carrier\_of \_ PO\_of ) \}.}
\end{description}
It has two components; a partial order and a proof that the carrier
set of the partial order is finite. Note that \texttt{FPO }is defined
as a dependent record type which inherits all the fields of \texttt{PO}.
Hence, 
\begin{itemize}
\item an object of type \texttt{FPO} can appear in any context where an
object of type \texttt{PO }is expected. 
\end{itemize}

\subsubsection*{Some more definitions }

For a finite partial order \texttt{P: FPO U }on some type \texttt{U}
let \texttt{C := Carrier\_of U P} and \texttt{R:= Rel\_of U P.} Then,
we have the following definitions:
\begin{description}
\item [{\texttt{Definition}}] \texttt{\emph{Is\_a\_chain\_in}}\texttt{
(e: Ensemble U): Prop:= }~\\
\texttt{(Included U e C /\textbackslash{} Inhabited U e)/\textbackslash{}
}~\\
\texttt{($\forall$ x y:U, (Included U (Couple U x y) e)-> R x y \textbackslash{}/
R y x). }
\end{description}
\begin{itemize}
\item Note that \texttt{\emph{Is\_a\_chain\_in}} is defined as a predicate.
It becomes true for any set \texttt{e: Ensemble U} iff \texttt{e }is
a non-empty set included in \texttt{C} which satisfies the chain condition. 
\item Also note that the first parameter in the above definition is a finite
partial order. We can avoid writing it using the \emph{section mechanism}
of Coq. 
\end{itemize}
Similarly we have, 
\begin{description}
\item [{\texttt{Definition}}] \texttt{\emph{Is\_an\_antichain\_in}}\texttt{
(e: Ensemble U): Prop := }~\\
\texttt{(Included U e C /\textbackslash{} Inhabited U e)/\textbackslash{}
}~\\
\texttt{($\forall$ x y:U, (Included U (Couple U x y) e)-> (R x y
\textbackslash{}/ R y x)}~\\
\texttt{-> x=y).}
\item [{\texttt{Inductive}}] \texttt{\emph{Is\_largest\_chain\_in}}\texttt{
(e: Ensemble U): Prop:= }~\\
\texttt{largest\_chain\_cond: }~\\
\texttt{Is\_a\_chain\_in e -> }~\\
\texttt{($\forall$ (e1: Ensemble U) (n n1:nat), }~\\
\texttt{Is\_a\_chain\_in e1 -> cardinal \_ e n -> cardinal \_ e1 n1
-> n1<= n)}~\\
\texttt{ -> Is\_largest\_chain\_in e.}
\end{description}
\begin{itemize}
\item In the above definition \texttt{cardinal} is a predicate which becomes
true for a set \texttt{S} and a natural number \texttt{n} iff \texttt{n}
is the size of the set \texttt{S}.
\end{itemize}
\begin{description}
\item [{\texttt{Inductive}}] \texttt{\emph{Is\_largest\_antichain\_in}}\texttt{
(e: Ensemble U): Prop:= }~\\
\texttt{largest\_antichain\_cond: }~\\
\texttt{Is\_an\_antichain\_in e -> }~\\
\texttt{($\forall$ (e1: Ensemble U) (n n1: nat), Is\_an\_antichain\_in
e1 -> }~\\
\texttt{cardinal \_ e n -> cardinal \_ e1 n1 -> n1<=n ) }~\\
\texttt{-> Is\_largest\_antichain\_in e. }
\item [{\texttt{Inductive}}] \texttt{\emph{Is\_a\_chain\_cover}}\texttt{
(cover: Ensemble (Ensemble U)): Prop:= }~\\
\texttt{cover\_cond: }~\\
\texttt{($\forall$ (e: Ensemble U), In \_ cover e -> Is\_a\_chain\_in
e)-> }~\\
\texttt{($\forall$ x:U, In \_ C x -> ($\exists$ e: Ensemble U, In
\_ cover e /\textbackslash{} In \_ e x)) -> Is\_a\_chain\_cover cover.}
\end{description}
\begin{itemize}
\item In the above definition \texttt{cover} is a collection of sets.
\end{itemize}
Here \texttt{\emph{Is\_a\_chain\_cover}} is a predicate which becomes
true for \texttt{cover} iff it satisfies the \texttt{cover\_cond}.
To satisfy the \texttt{cover\_cond} every member of \texttt{cover}
should be a chain and every element of the carrier set \texttt{C }must
be covered by some chain in the \texttt{cover}. Similarly we have, 
\begin{description}
\item [{\texttt{Inductive}}] \texttt{\emph{Is\_an\_antichain\_cover}}\texttt{
(cover: Ensemble (Ensemble U)): Prop:=}~\\
\texttt{ AC\_cover\_cond: }~\\
\texttt{($\forall$ (e: Ensemble U), In \_ cover e -> Is\_an\_antichain\_in
e)-> ($\forall$ x:U, In \_ C x -> ($\exists$ e: Ensemble U, In \_
cover e /\textbackslash{} In \_ e x)) }~\\
\texttt{-> Is\_an\_antichain\_cover cover.}
\item [{\texttt{Inductive}}] \texttt{\emph{Is\_a\_smallest\_chain\_cover}}\texttt{
(scover: Ensemble (Ensemble U)): Prop:= }~\\
\texttt{smallest\_cover\_cond: }~\\
\texttt{(Is\_a\_chain\_cover P scover) -> }~\\
\texttt{( $\forall$ (cover: Ensemble (Ensemble U))(sn n: nat), }~\\
\texttt{(Is\_a\_chain\_cover P cover /\textbackslash{} cardinal \_
scover sn /\textbackslash{} cardinal \_ cover n) -> (sn <=n) ) }~\\
\texttt{-> Is\_a\_smallest\_chain\_cover P scover.}
\item [{\texttt{Inductive}}] \texttt{\emph{Is\_a\_smallest\_antichain\_cover}}\texttt{
(scover: Ensemble (Ensemble U)): Prop:= }~\\
\texttt{smallest\_cover\_cond\_AC: }~\\
\texttt{(Is\_an\_antichain\_cover P scover) -> }~\\
\texttt{( $\forall$ (cover: Ensemble (Ensemble U))(sn n: nat), }~\\
\texttt{(Is\_an\_antichain\_cover P cover /\textbackslash{} cardinal
\_ scover sn /\textbackslash{} }~\\
\texttt{cardinal \_ cover n) -> (sn <=n) ) }~\\
\texttt{-> Is\_a\_smallest\_antichain\_cover P scover.}
\item [{\texttt{Inductive}}] \texttt{\emph{Is\_height }}\texttt{(n: nat)
: Prop:= }~\\
\texttt{H\_cond: }~\\
\texttt{($\exists$ lc: Ensemble U, Is\_largest\_chain\_in P lc /\textbackslash{}
cardinal \_ lc n) -> (Is\_height P n). }
\item [{\texttt{Inductive}}] \texttt{\emph{Is\_width}}\texttt{ (n: nat)
:Prop := }~\\
\texttt{W\_cond: }~\\
\texttt{($\exists$ la: Ensemble U, Is\_largest\_antichain\_in P la
/\textbackslash{} }~\\
\texttt{cardinal \_ la n)-> (Is\_width P n).}
\end{description}

\subsection{Theorem statements}

The definitions, we have seen so far, are sufficient to express the
formal statement of Dilworth's Theorem in Coq. 
\begin{description}
\item [{\texttt{Theorem}}] \texttt{Dilworth: $\forall$ (P: FPO U), Dilworth\_statement
P.}
\end{description}
where \texttt{Dilworth\_statement} is defined as, 
\begin{description}
\item [{\texttt{Definition}}] \texttt{Dilworth\_statement:= }~\\
\texttt{fun (P: FPO U)=> }~\\
\texttt{$\forall$ (m n: nat), (Is\_width P m) -> }~\\
\texttt{($\exists$ cover: Ensemble (Ensemble U), (Is\_a\_smallest\_chain\_cover
P cover)/\textbackslash{} (cardinal \_ cover n)) }~\\
\texttt{-> m=n.}
\end{description}
At this point one can easily verify that the combined meaning of all
the terms in the above statement corresponds to the actual Dilworth's
Theorem. Similarly for dual-Dilworth we proved, 
\begin{description}
\item [{\texttt{Theorem}}] \texttt{Dual\_Dilworth: $\forall$ (P: FPO U),
Dual\_Dilworth\_statement P.}
\end{description}
where, \texttt{Dual\_Dilworth\_statement} is defined as,
\begin{description}
\item [{\texttt{Definition}}] \texttt{Dual\_Dilworth\_statement:=}~\\
\texttt{ fun (P: FPO U)=> }~\\
\texttt{$\forall$ (m n: nat), (Is\_height P m) -> }~\\
\texttt{($\exists$ cover: Ensemble (Ensemble U), (Is\_a\_smallest\_antichain\_cover
P cover)/\textbackslash{} (cardinal \_ cover n))}~\\
\texttt{ -> m=n.}
\end{description}
The original proofs of these theorems are classical in nature, hence,
we need the principal of excluded middle at many places. At certain
points, we also need to extract functions from relations. Therefore,
we imported the \emph{Classical} and \emph{ClassicalChoice} modules
of the standard library, which assumes the following three axioms:
\begin{description}
\item [{\texttt{Axiom}}] \texttt{classic : $\forall$ P:Prop, P \textbackslash{}/
\textasciitilde{} P.}
\item [{\texttt{Axiom}}] \texttt{dependent\_unique\_choice : }~\\
\texttt{$\forall$ (A:Type) (B:A -> Type) (R:$\forall$ x:A, B x ->
Prop), }~\\
\texttt{($\forall$ x : A, $\exists$! y : B x, R x y) ->}~\\
\texttt{($\exists$ f : ($\forall$ x:A, B x), $\forall$ x:A, R x
(f x)).}
\item [{\texttt{Axiom}}] \texttt{relational\_choice : $\forall$ (A B : Type)
(R : A->B->Prop), }~\\
\texttt{($\forall$ x : A, $\exists$ y : B, R x y) -> }~\\
\texttt{$\exists$ R' : A->B->Prop, subrelation R' R /\textbackslash{}
$\forall$ x : A, $\exists$! y : B, R' x y.}
\end{description}

\section{Some results on finite partial orders\label{sec:Some-results-on-FPO}}

In this section we explain some results of general nature on finite
partial orders. These results are used at more than one places in
the proof of Dilworth's Theorem. They are proved as Lemmas and compiled
in a separate file. For most of the Lemmas their statements can be
inferred from their names. Here, we only provide an English language
description of some of them. 

\subsection*{Existence proofs}

A large number of lemmas are concerned with the existence of a defined
object. For example, in our proof when we say ``Let A be an antichain
of the poset P...'' we assume that there exists an antichain for
the poset P. However, in a formal system like Coq, we need a proof
of existence of such object before we can instantiate it. Following
is a partial list of such results: 
\begin{description}
\item [{Lemma-1}] \emph{Chain\_exists}: There exists a chain in every finite
partial order (FPO). \textbf{Proof.} Trivial.
\item [{Lemma-2}] \emph{Chain\_cover\_exists:} There exists a chain cover
for every FPO.\\
\textbf{Proof.} Trivial.
\item [{Lemma-3}] \emph{Minimal\_element\_exists:} The set minimal(P) is
non-empty for every P: FPO. \textbf{Proof.} Using induction on the
size of P.
\item [{Lemma-4}] \emph{Maximal\_element\_exists:} The set maximal(P) is
non-empty for every P: FPO. \textbf{Proof.} Using induction on the
size of P.
\item [{Lemma-5}] \emph{Largest\_element\_exists:} If a finite partial
order is also totally ordered then there exists a largest element
in it. \textbf{Proof.}\textbf{\emph{ }}The maximal element becomes
the largest element and we know that there exists a maximal element. 
\item [{Lemma-6}] \emph{Minimal\_for\_every\_y:} For every element $y$
of a finite partial order P there exists an element $x$ in P such
that $x\leq y$ and $x\in\text{minimal(P)}$. \emph{}\\
\textbf{Proof.} Let $X=\{x:P|\,x\le y\}$. Then the poset $(X,\leq)$
will have a minimal element, say $x_{0}$. It is also a minimal element
of P. 
\item [{Lemma-7}] \emph{Maximal\_for\_every\_x:} For every element $x$
of a finite partial order P there exists an element $y$ in P such
that $x\leq y$ and $y\in\text{maximal(P)}$. \emph{}\\
\textbf{Proof.} Let $Y=\{y:P|\,x\le y\}$. Then the poset $(Y,\leq)$
will have a maximal element, say $y_{m}$. It is also a maximal element
of P. 
\item [{Lemma-8}] \emph{Largest\_set\_exists:} There exists a largest set
(by cardinality) in a finite and non-empty collection of finite sets.\textbf{\emph{
}}\textbf{Proof. }Consider the collection of sets together with the
strict set-inclusion relation. This forms a finite partial order.
Any maximal element of this finite partial order will be a largest
set. Moreover, such a maximal element exists due to Lemma-4.
\item [{Lemma-9}] \emph{exists\_largest\_antichain:} In every finite partial
order there exists a largest antichain.\emph{ }\textbf{Proof.} Note
that this statement is not true for partial orders. The proof is similar
to Lemma-7. 
\item [{Lemma-10}] \emph{exists\_largest\_chain:} In every finite partial
order there exists a largest antichain.\emph{ }\textbf{Proof.} Again,
its true only for finite partial orders. Proof is similar to Lemma-7.
\end{description}

\subsection*{Some other proofs}

When dealing with sets the set-inclusion relations occurs more naturally
than the comparison based on the set sizes. Therefore, we defined
a binary relations \emph{Inside} ($\prec$) on the collection of all
the finite partial orders. 
\begin{itemize}
\item We say $P_{1}\prec P_{2}$ iff carrier set of $P_{1}$ is strictly
included in the carrier set of $P_{2}$ and both the posets are defined
on the same binary relation. 
\end{itemize}
Then to use well-founded induction we proved that the relation $\prec$
is well founded. 
\begin{description}
\item [{Lemma-11}] \emph{Inside\_is\_WF:} The binary relation Inside (i.e,
$\prec$ ) is well founded on the set of all finite partial orders.
\textbf{Proof.} Using strong induction on the size of finite partial
orders. 
\item [{Lemma-12}] \emph{Largest\_antichain\_remains:} If $\mathcal{A}$
is a largest antichain of $P_{2}$ and $P_{1}\prec P_{2}$ then $\mathcal{A}$
is also a largest antichain in $P_{1}$ provided $\mathcal{A}\subset P_{1}$.
\textbf{}\\
\textbf{Proof. }Assume otherwise, then there will be a larger antichain
say $\mathcal{A}'$ in $P_{1}$. This will also be larger in $P_{2}$,
which contradicts.
\item [{Lemma-13}] \emph{NoTwoCommon:} A chain and an antichain can have
at most one element in common. \textbf{Proof.} Trivial.
\item [{Lemma-14}] \emph{Minimal\_is\_antichain:} Minimal(P) is an antichain
in P. \textbf{}\\
\textbf{Proof.} Trivial.
\item [{Lemma-15}] \emph{Maximal\_is\_antichain:} Maximal(P) is an antichain
in P. \textbf{}\\
\textbf{Proof. }Trivial. 
\item [{Lemma-16}] \emph{exists\_disjoint\_cover:} If $\mathcal{C_{V}}$
is a smallest chain cover of size $m$ for P, then there also exists
a disjoint chain cover $\mathcal{C_{V}}'$ of size $m$ for P. \textbf{}\\
\textbf{Proof.} Using induction on $m$.
\end{description}

\subsection*{Wrapping Up}

This work is done in the Coq Proof General (Version 4.4pre). We have
used the Company-Coq extension \cite{key-14-1} for the Proof General.
The proof is split into the the following nine files:
\begin{enumerate}
\item \texttt{PigeonHole.v}: It contains some variants of the Pigeonhole
Principal.
\item \texttt{BasicFacts.v}: Contains some useful properties on numbers
and sets. It also contains strong induction and some variants of Choice
theorem. 
\item BasicFacts$_{2}$.v: Contains some more facts about power-sets and
binomial coefficients. 
\item \texttt{FPO\_Facts.v}: Most of the definitions and some results on
finite partial orders are proved in this file. 
\item \texttt{FPO\_Facts$_{2}$.v}: Contains most of the lemmas that we
discussed in this section. 
\item \texttt{FiniteDilworth\_AB.v}: Contains the proofs of forward and
backward directions of Dilworth's Theorem.
\item \texttt{FiniteDilworth.v}: Contains the proof of the main statement
of Dilworth's Theorem. 
\item \texttt{Dual\_Dilworth.v}: Contains the proof of the Dual-Dilworth
Theorem. 
\item \texttt{Combi\_1.v}: Some new tactics are defined to automate the
proofs of some trivial facts on numbers, logic, sets and finite partial
orders. 
\end{enumerate}
The Coq code for this work is available at \cite{key-15}. The files
can be safely compiled in the given order. 

\section{Related Work\label{sec:Related-Work}}

Rudnicki \cite{key-9} presents a formalization of Dilworth's Theorem
in Mizar. In the same paper they also provide a proof of Erd\H{o}s-Szekeres
Theorem \cite{key-6-4} using Dilworth's Theorem. A separate proof
of the Hall's marriage Theorem in Mizar appeared in \cite{key-10}.
Jiang and Nipkov \cite{key-11} also presented two different proofs
of Hall's Theorem in HOL/Isabella. All these formalizations stand
isolated because they are built on different platforms using different
libraries of facts and definitions. Our work aims to create a homogeneous
set of definitions that can be used efficiently in the formalization
of all these and many other important results. Moreover, we selected
a different theorem prover, the Coq Proof Assistant, to reduce the
time and effort of formalization.

\section{Conclusions\label{sec:Conclusions}}

Formalization of any mathematical theory is a difficult task. It requires
a lot of time and effort. In combinatorics, the task becomes even
more difficult due to the lack of structure in the theory. Some statements
often admit more than one proof using completely different ideas.
There is no predefined order to mechanize the theory. Any such effort
would require to explore the dependencies among the results and identify
an effective order to formalize them. Our work aims in this direction and 
the main contributions of this paper are: 
\begin{enumerate}
\item Fully formalized proofs of Dilworth's and Mirsky's decomposition theorems
in Coq, together with a detailed account of all the definitions and
the theorem statement. 
\item A clear compilation of more general results and definitions which
could be useful in other similar formalizations. 
\end{enumerate}
The Coq code for this work is accessible online at \cite{key-15}.
In the immediate future, one can use this work to formalize other
related results like Hall's Theorem \cite{key-6-1,key-6-2}, Erd\H{o}s-Szekeres
Theorem \cite{key-6-4} and Sperner's Lemma \cite{key-6-5,key-6-3}.
One can also attempt to formalize the infinite version of Dilworth's
Theorem. 

In the long run we would like to see most of the well known results
from this field formalized and organized in the form a library. This
would significantly reduce the time of formalization of any new result
from this area. 

\subsection*{Acknowledgements }

We are thankful to Mohit Garg for explaining these proofs and motivating
us to formalize them.

\end{document}